\documentclass{ragtime} 

\title[Numerical simulations of thin accretion discs with PLUTO]%
      {Numerical simulations of thin accretion discs with PLUTO}

\author[V. Parthasarathy, 
        W. Klu\'{z}niak    
        ]
       {Varadarajan Parthasarathy\at{1,a} 
        W{\l}odzimierz Klu\'{z}niak\at[]{1}\\ 
        \ins{1}Copernicus Astronomical Center, ul. Bartycka 18, 00-716 Warszawa, Poland \splitins[1]
        \\
        \ins{a}\Email{varada@camk.edu.pl}} 

\coentry{P. Varadarajan, W. Kluzniak}%
       {Numerical simulation of thin disc with PLUTO}

\begin{document}

\begin{abstract}
  Our goal is to perform global simulations of thin accretion discs around compact bodies like neutron stars with dipolar
  magnetic profile and black holes by exploiting the facilities provided by state-of-the-art grid-based, high 
  resolution shock capturing (HRSC) and finite volume codes. We have used the Godunov-type code PLUTO to simulate a
  thin disc around a compact object prescribed with a pseudo-Newtonian potential in a purely hydrodynamical (HD) regime,
  with numerical viscosity as a first step towards achieving our goal as mentioned above. 
\end{abstract}

\begin{keywords}
Accretion discs~-- pseudo--Newtonian~-- PLUTO~-- hydrodynamics~-- compact objects ~-- black holes
\end{keywords}

\section{INTRODUCTION}\label{intro}

Disc like structures are ubiquitous as known from several astronomical observations. It is now understood that these are the 
result of accretion flows, which have been studied by the astrophysical community since 1968. The first analytic solution was 
obtained by \cite{1973A&A....24..337S}, hereafter SS,
preceded by a numerical solution obtained by \cite{1968ApJ...151L..83P}. The model of SS was geometrically thin
steady accretion discs and since 1973, their approach has become a standard framework, which assumes that irrespective
of the physics involved in the production of stress, the result scales with the pressure. The main features of the SS model are the $\alpha$ viscosity 
prescription and the assumption of vertical extent of the disc being smaller than its radial scale. This gives rise to a small parameter or disc-aspect ratio
$\epsilon \equiv c_{s}/\Omega r$, where $c_{s}$ is the sound speed and $\Omega$ is the Keplerian angular velocity, which allows detailed solutions for the flow \citep{2000astro.ph..6266K, 2002A&A...396..623R}.
The origin of viscosity in accretion discs and the exact mechanism of angular momentum transport is still not understood with clarity, 
since $\alpha$ prescription is valid for vertically averaged thin discs, however it is worth noting that \cite{1991ApJ...376..214B} have proposed the magneto-rotational 
instability (MRI) as the origin of MHD turbulence, which is efficient in transporting angular momentum. There is a consensus that MRI is the origin of viscosity in accretion discs.

Our motivation to perform numerical simulations is to determine an appropriate model in three dimensional time-evolution scenario, which incorporates known physical ideas
along with a robust numerical scheme.

\section{NUMERICAL CODE: PLUTO}\label{code}

PLUTO \citep{2007ApJS..170..228M} is a Godunov-type shock-capturing code, constructed to integrate system of conservation laws given as
\begin{equation}
\frac{\partial \textbf{U}}{\partial t} = - \nabla \cdot \textbf{T}(\textbf{U}) + \textbf{S}(\textbf{U}) ,      \label{pl1}
\end{equation}
where $\textbf{U}$ represents a state vector of conserved quantities, $\textbf{T}(\textbf{U})$ represents fluxes of each component of state vector
and $\textbf{S}(\textbf{U})$ defines the source terms. PLUTO provides a modular environment capable of simulating hypersonic flows in presence of 
discontinuities in multi-dimensional Cartesian and curvilinear coordinates. The code in its current version (v4.0) is equipped with four independent physics
modules, namely hydrodynamics (HD), magnetohydrodynamics (MHD), relativistic hydrodynamics (RHD) and relativistic magnetohydrodynamics (MHD), which perform 
numerical integration of the Euler/Navier-Stokes equations, ideal/resistive MHD equations, energy-momentum conservation laws of special relativistic perfect gas,
and equations of special relativistic magnetized ideal plasma.

In the HD module we numerically solve the following equations:
\begin{gather}
 \frac{\partial \rho}{\partial t} + \nabla \cdot (\rho \textbf{v}) = 0 \\        
 \frac{\partial \rho \textbf{v}}{\partial t} + \nabla \cdot (\rho \textbf{v} \textbf{v} + p \textbf{I}) = -\rho \nabla \Phi \\   
 \frac{\partial E}{\partial t} + \nabla \cdot [ (E + p) \textbf{v} ] = \rho \textbf{v} \cdot \textbf{g}
\end{gather}
where the conservative variables, fluxes and source terms are
\begin{equation}
\textbf{U} = \left( \begin{array}{c}                                                   
                         \rho \\
                         \textbf{m} \\
                         E \end{array} \right)\qquad \text{,} \qquad
\textbf{T}(\textbf{U}) = \left( \begin{array}{c}
                                     \rho \textbf{v}^{T} \\
                                     \rho \textbf{v} \textbf{v} + p \textbf{I} \\
                                     (E + p) \textbf{v}^{T} \end{array} \right) \qquad \text{,} \qquad
\textbf{S}(\textbf{U}) = \left( \begin{array}{c}
                                 0 \\
                                 -\rho \nabla \Phi \\
                                 \rho \textbf{v} \cdot \textbf{g} \end{array} \right) .
\end{equation}
The mass density is $\rho$, momentum density is $\textbf{m} = \rho \textbf{v}$, pressure is $p$, acceleration vector is $\textbf{g}$ and the total energy density $E$ is
\begin{equation}
 E = \rho \epsilon + \frac{\textbf{m}^{2}}{2 \rho}
\end{equation}
where an equation of state provides the closure $p = p(\rho, \epsilon)$. For a polytrope \footnote{In this code adiabatic index  is same as polytropic index, hence $\gamma = \Gamma$.}, 
with $\gamma = 5/3$, the total energy density is
\begin{equation}
 E = \frac{p}{\gamma - 1} + \frac{\textbf{m}^{2}}{2 \rho} .
\end{equation}

In PLUTO, the numerical integration of Eq.(1) is performed with high-resolution shock-capturing scheme (HRSC), where the algorithm employed will capture the 
discontinuities in the solution and smear them over few grid cells, without producing spurious oscillations near the discontinuities, using finite volume 
methods in which discrete data is represented as averages over a control volume on structured grids. Generically, the paradigm of HRSC was developed by 
merging Godunov-type methods with advanced numerical methods, capable of obtaining higher order accuracy in smooth parts of the solution and provide 
higher resolution of discontinuities without large smearing over grids.

\section{NUMERICAL SIMULATIONS} \label{nusim}

\subsection{Initial Conditions}
We perform simulations in spherical coordinates $(R, \theta)$, in 2.5 dimensions (2.5D) assuming axisymmetry around the rotation axis of the disc. 2.5D considers two
spatially independent coordinates, but all three components of velocities (also magnetic fields if present). The setup \citep{2009A&A...508.1117Z} consists of
a thin disc, a corona and a compact body at the center whose gravitational potential \citep{2002MNRAS.335L..29K} we take to be 
\begin{equation}
 \Phi(R) = -\frac{1}{6} \exp\left(\frac{6r_{g}}{R} - 1 \right)
\end{equation}
where gravitational radius $r_{g} = 1$. The initial density and thermal pressure of the disc are determined by the vertical hydrostatic equilibrium
\begin{gather}
 \rho_{d} = \left( \frac{2}{5 \epsilon^{2}} \left[\frac{1}{R} - \left(1 - \frac{5 \epsilon^{2}}{2}\right)\frac{1}{r}\right]\right)^{3/2} \\
 p_{d} = \epsilon^{2} \rho_{d}^{5/3}
\end{gather}
where cylindrical radius is given as $r = R$ sin$(\theta)$, $\gamma = 5/3$, $\epsilon = 0.1$. The azimuthal velocity is obtained from the radial equilibrium
\begin{equation}
 v_{\phi d} = \sqrt{ \frac{\exp(\frac{6}{r} - 1)}{r} }
\end{equation}
and the meridional flow is given as
\begin{equation}
 v_{Rd} = -\alpha \epsilon^{2} \left[  10 - \frac{32}{3} \Lambda \alpha^{2} - \Lambda \left( 5 - \frac{1}{\epsilon^{2} \tan^{2}\theta} \right) \right] \sqrt{\frac{1}{R \sin^{3}\theta}}
\end{equation}
with $\alpha = 0.01$ and $\Lambda = \frac{11}{5} / \left( 1 + \frac{64}{25} \alpha^{2} \right)$ \citep{2000astro.ph..6266K}. The corona is a non-rotating polytrope, with density and pressure given as
\begin{gather}
 \rho_{c} = \rho_{a} \left( \frac{1}{R} \right)^{\frac{1}{\gamma -1}} \\
  p_{c} = \frac{\gamma -1}{\gamma} \left( \frac{1}{R} \right)^{\frac{\gamma}{\gamma -1}} .
\end{gather}
The density contrast between corona and disc is set by the parameter $\rho_{a} = 0.01$. 

\subsection{Computational Domain and Boundary Conditions}

The computational domain is a two dimensional box in $(R, \theta)$, angular coordinate spanning the sector  $[0,\pi/2]$ and delimited by the radial coordinate extending over
$[3,20]$. Both the radial and angular coordinates are discretized with 127 points on a uniform grid. As we perform 2.5D simulation there is axisymmetry about the rotation axis 
and planar symmetry with respect to the disc mid-plane. At the inner edge of the disc, we employ outflow (zero gradient) boundary condition and at the outer edge of the disc we 
prescribe a numerical condition, such that the flow from the inner boundary is fed back to the outer boundary, which aids in conserving the mass by preventing the disc
from being drained. 

\subsection{A Note On Viscosity}

In the simulation reported here we have not employed a physical prescription for viscosity, which in PLUTO is achieved by adding viscous stress tensor to Eq.(1) 
\begin{equation}
 \frac{\partial \textbf{U}}{\partial t} + \nabla \cdot \textbf{T} = \nabla \cdot \Pi + \textbf{S}
\end{equation}
where $\Pi$ is the viscous stress tensor, with components
\begin{equation}
 (\Pi)_{ij} = 2 \frac{\eta}{h_{i}h_{j}} \left( \frac{v_{i;j} + v_{j;i}}{2} \right) + \left( \eta_{b} - \frac{2}{3} \eta \right) \nabla \cdot \textbf{v}\delta_{ij}
\end{equation}
where $\eta$, $\eta_{b}$ are shear and bulk coefficient of viscosity, $v_{i;j}$, $v_{j;i}$ are covariant derivatives, and $h_{i}$, $h_{j}$ are geometrical elements in the corresponding
directions respectively. 

The most conspicuous phenomenon while performing numerical simulations of the equations of fluid dynamics is shock waves, which physically is a transition zone across which
ram pressure is converted into thermal pressure and kinetic energy into enthalpy. Numerical treatment of shocks\footnote{Personal communication with numerical experts.} is a 
complicated issue, which was dealt with by adding large but nonphysical value of viscosity to the algorithms such that the narrow transition zones got thickened and it was possible to handle 
shocks computationally. This is known as artificial viscosity, introduced for numerical purposes for ease in computational treatment of physical processes. It is to be noted that artificial
viscosity is different from numerical viscosity which is a result of smoothing effect. 

Convective flux exchanges momentum between neighboring elements and the resultant in a given element is then added to the existing momentum in order to get an average for that element. 
As the time step advances the previously calculated average value is passed to the next element and consecutive steps of such smoothing effect will create a diffusion of momentum along the flow. 
However such a numerical diffusion which depends only on fluid convection, does not work like viscous stress as shown in Eq.(15) which satisfies well known physical laws. We still exploit 
numerical viscosity for our simulation by simply trusting the robustness of the approximate numerical schemes in PLUTO, that has been tested against several benchmark test problems.

\subsection{Discussion of Simulations}

Our simulation is intended to be a substrate on which we plan to work further and not a scientific result at present. We perform the simulation for about 100 orbital periods corresponding to the innermost
stable circular orbit. We observed that the part of the disc within the innermost stable circular orbit was not stable and in a few rotations the disc relaxed by shedding some of its mass, which
through our boundary condition was fed back to the disc, thereby preventing the disc from being drained. As expected the disc finally reaches stationary state with the inner edge at the innermost stable 
circular orbit ($r=6$). As mentioned previously we have used numerical viscosity, which is responsible for transporting angular momentum outwards. We have delimited the angular sector such that the radial
grid starts three units from the origin due to numerical reasons. While testing our routines further, we are able to overcome this issue by physical prescription of viscosity, however due to 
constraints in time we are yet to perform simulations with physical viscosity.

We present the results obtained from our simulation as follows. Fig. (1) shows the initial appearance of the disc and corona described by Eq. (9) \& Eq. (13). The evolution of the disc into a steady
state following a relaxation process is shown in Fig. (2) \& Fig. (4) and the distribution of density at the midplane of the disc is plotted in Fig. (3) \& Fig. (5) respectively. The profiles of the azimuthal velocity
at the midplane of the disc are plotted in Fig. (6) \& Fig. (7), comparison of these two figures shows the stability of the velocity profile at $t\geq$ 10.
\begin{figure}
  \begin{center}
 \includegraphics[width=\textwidth]{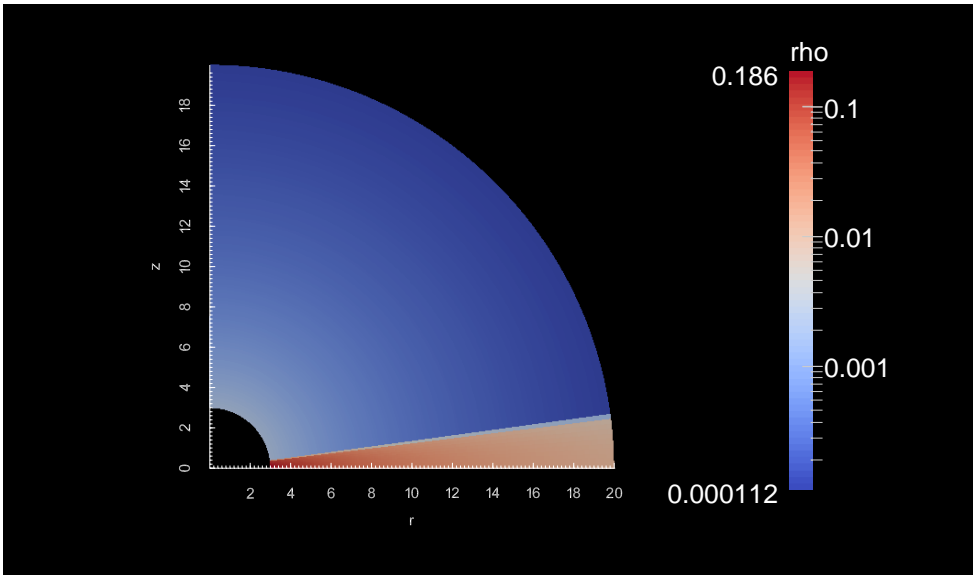}
 \end{center}
 \caption{\label{Fig:1} The initial appearance of disc and corona. Colors represent logarithmic density.}
\end{figure}

\begin{figure}
 \begin{center}
 \includegraphics[width=\textwidth]{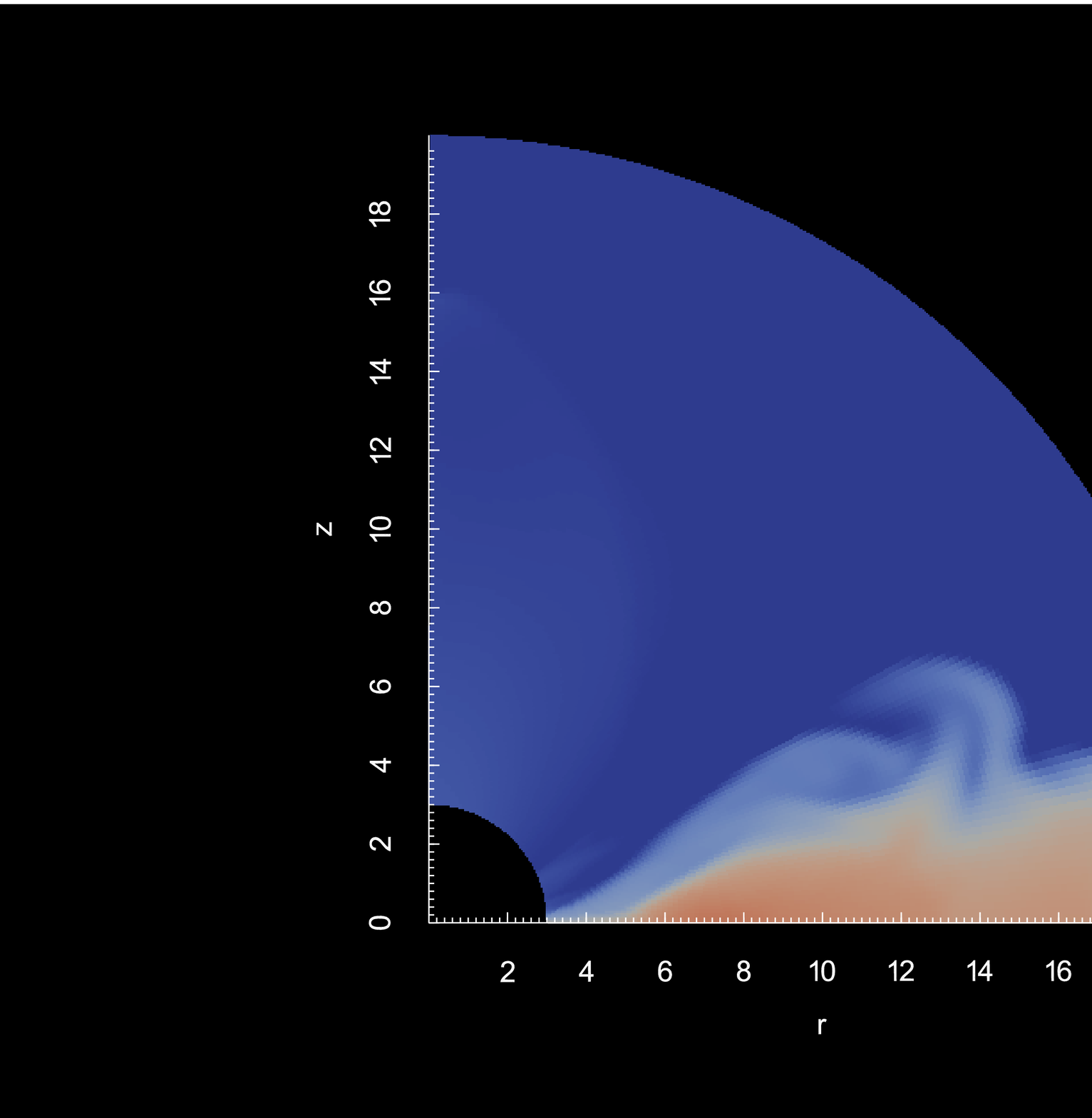}
 \caption{\label{Fig:2} Logarithmic plot of density at $t=10.0$.}
 \end{center}
\end{figure}

\begin{figure}
 \begin{center}
 \includegraphics[width=0.5\textwidth]{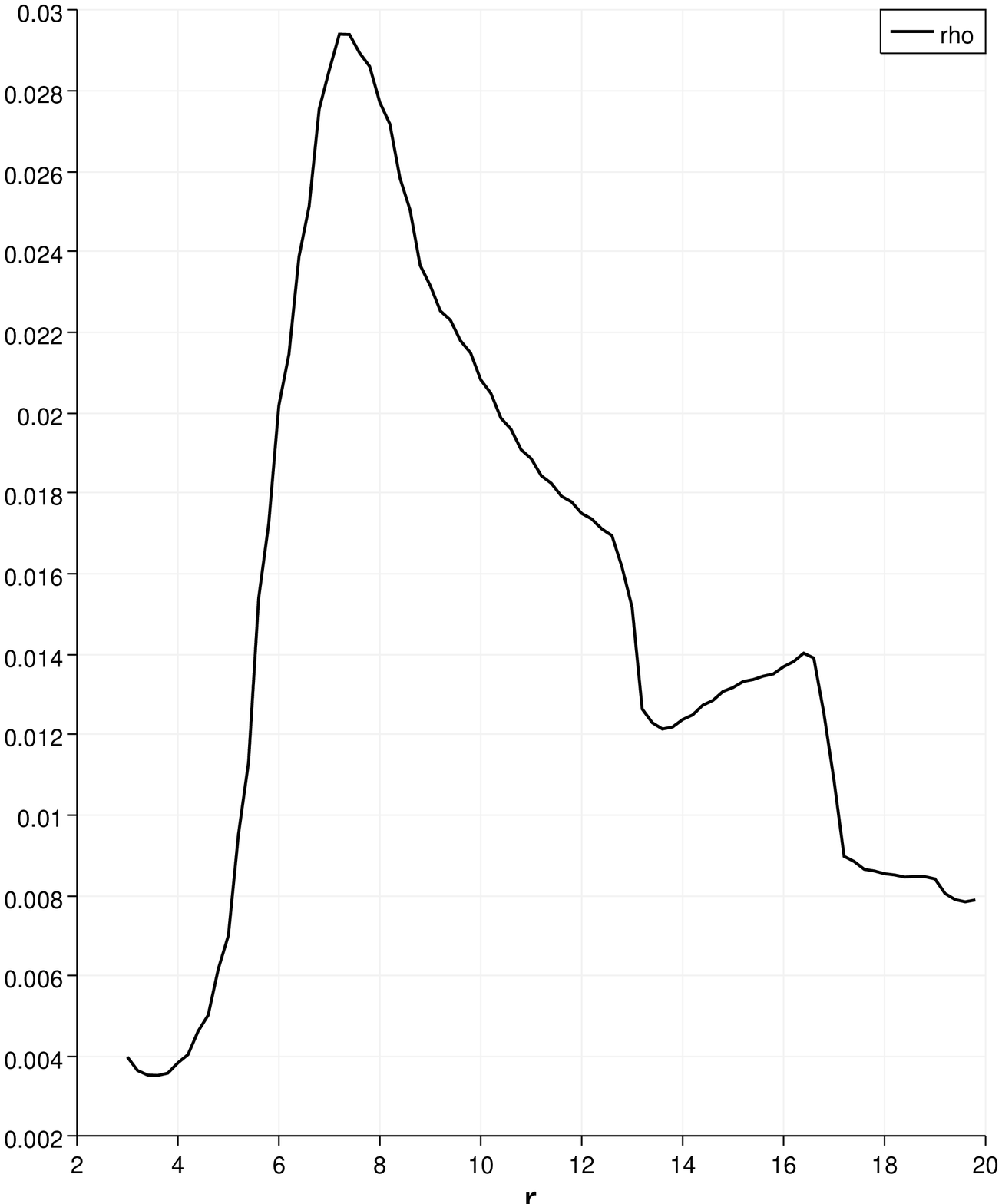}
 \caption{\label{Fig:3} Density in the midplane of disc at $t=10.0$.}
 \end{center}
\end{figure}

\begin{figure}
 \begin{center}
 \includegraphics[width=\textwidth]{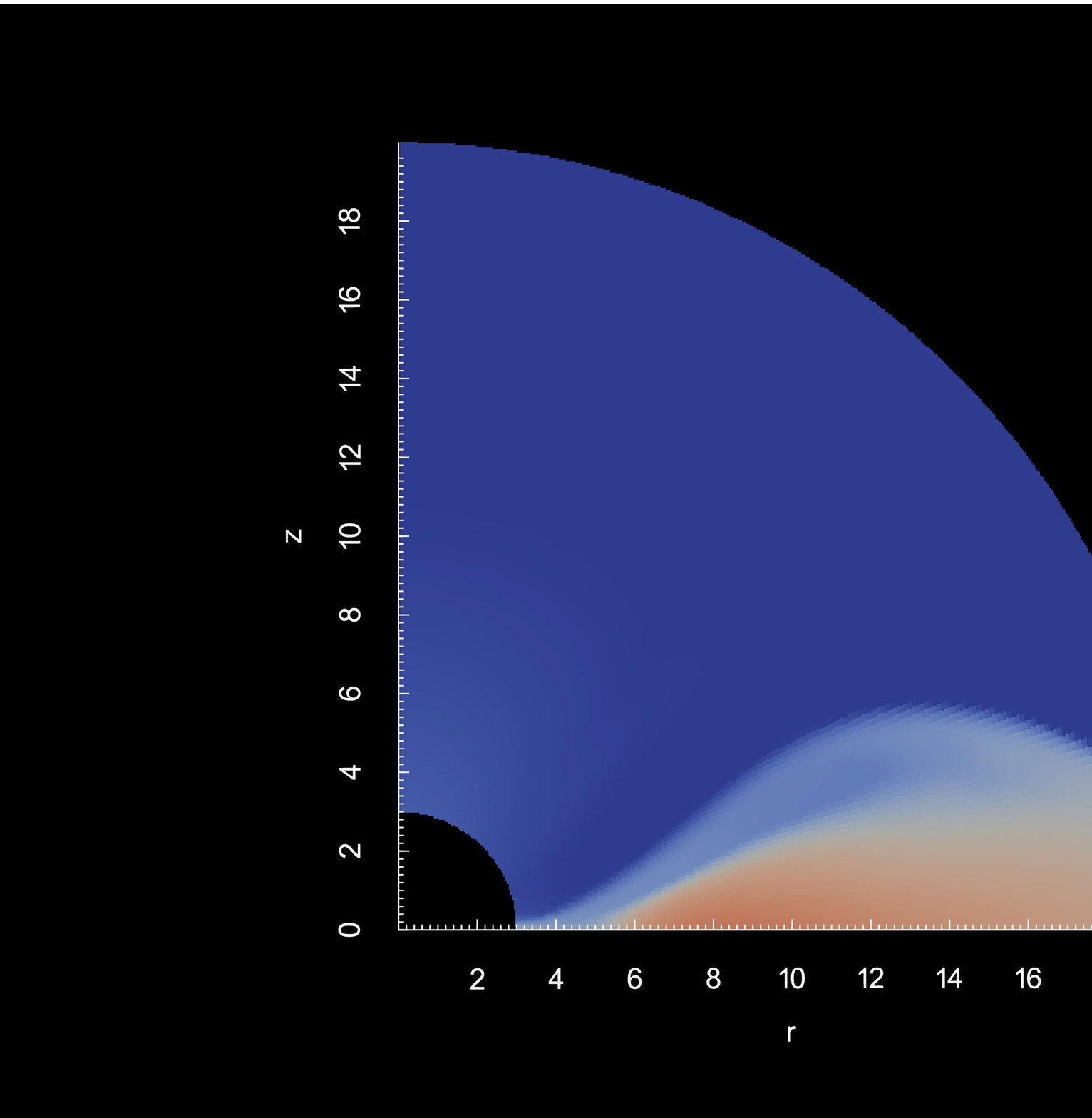}
 \caption{\label{Fig:4} Logarithmic plot of density at $t=70.0$.}
 \end{center}
\end{figure}

\begin{figure}
 \begin{center}
 \includegraphics[width=0.5\textwidth]{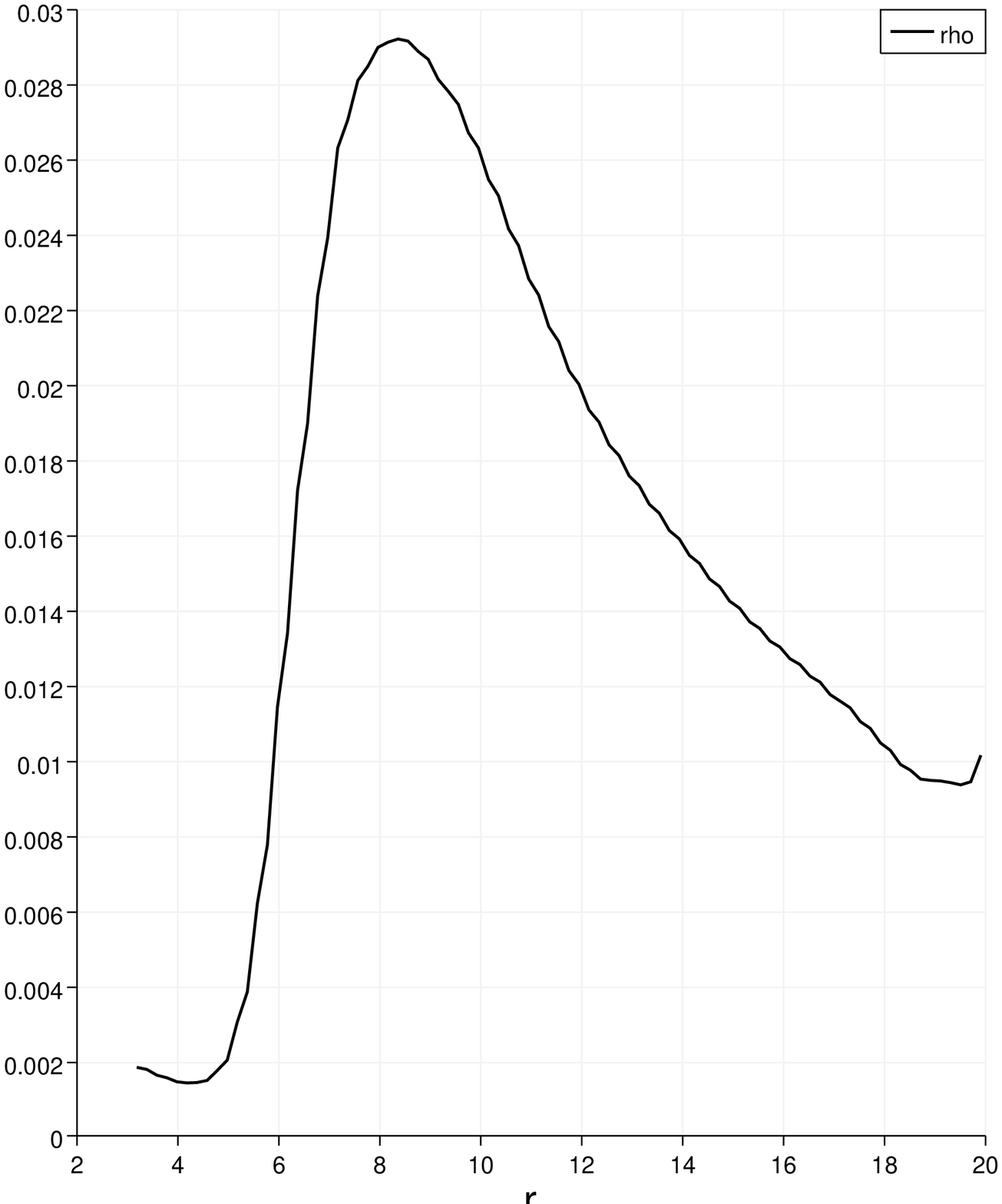}
 \caption{\label{Fig:5} Density in the midplane of disc at $t=70.0$.}
 \end{center}
\end{figure}

\begin{figure}
 \begin{center}
 \includegraphics[width=0.5\textwidth]{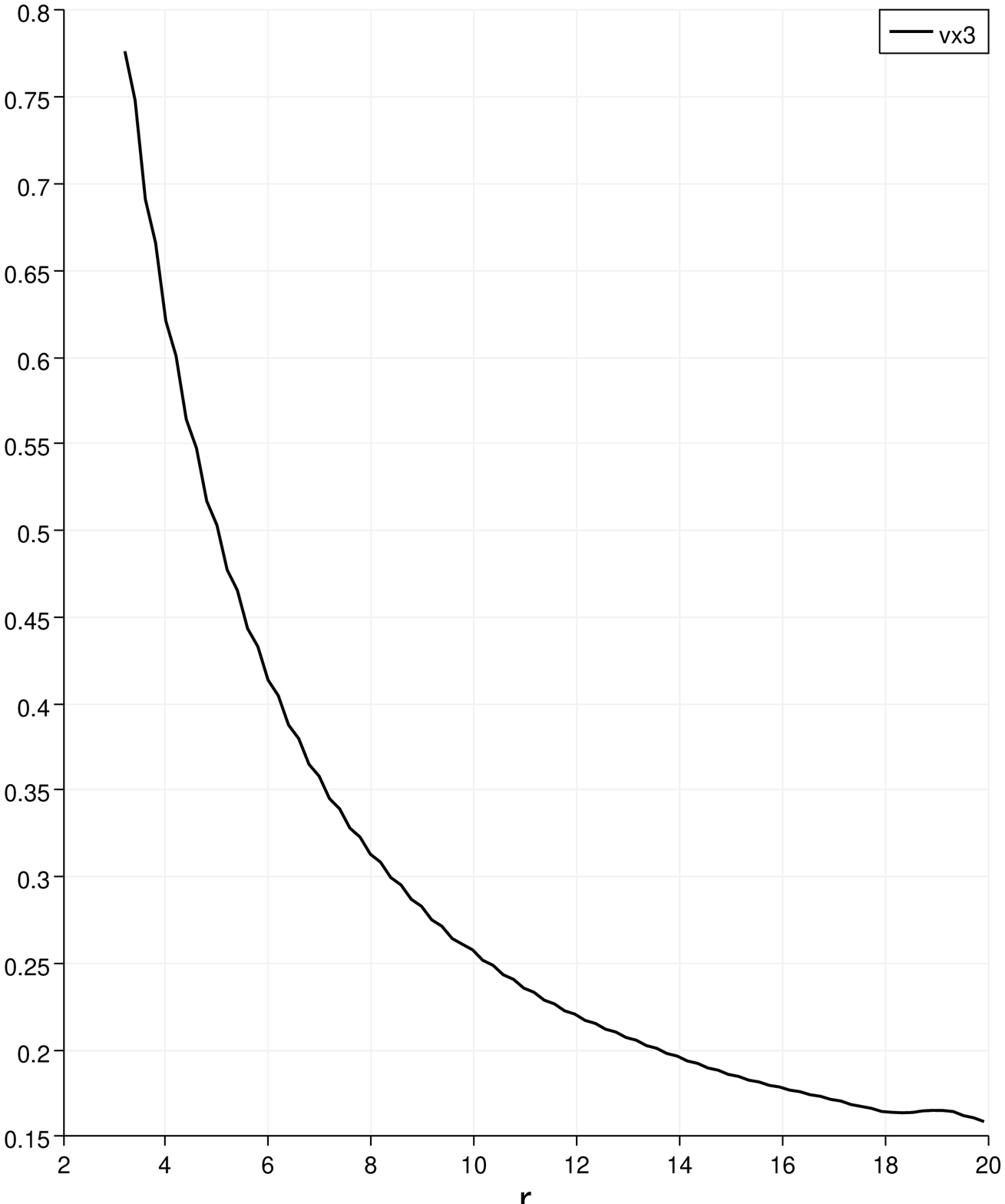}
 \caption{\label{Fig:2} Keplerian profile of azimuthal velocity in the midplane of disc at $t=10.0$.}
 \end{center}
\end{figure}

\begin{figure}
 \begin{center}
 \includegraphics[width=0.5\textwidth]{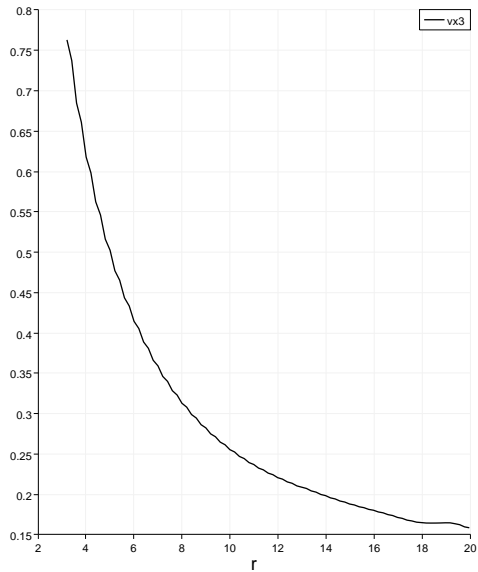}
 \caption{\label{Fig:2} Keplerian profile of azimuthal velocity in the midplane of disc at $t=70.0$.}
 \end{center}
\end{figure}

\newpage

\section{CONCLUSIONS}

We have performed numerical simulations using the hydrodynamical module in PLUTO to obtain steady thin discs around a compact object prescribed with a pseudo-Newtonian potential.
The disc relaxes in few rotations and reaches a steady state with its inner edge at innermost stable circular orbit. The profile of the azimuthal velocity remains Keplerian 
throughout. We plan to improve our routines in the code and successfully perform global simulations of thin discs around neutron stars with dipolar magnetic field. 


\ack{We would like to acknowledge Dr. Miljenko {\v C}emelji\'c for his crucial guidance and interesting discussions with nuances of the code.}
This work was supported by NCN grant 2013/08/A/ST9/00795.

\bibliography{partha}
\end{document}